\documentstyle[12pt]{article}

\begin{document}
\begin{titlepage}
\begin{flushright}
ILL-(TH)-94-\#23~~~~~~~~~~~\\
NOVEMBER 1994   ~~~~~~~~~~~\\
HLRZ-60/94      ~~~~~~~~~~~
\vskip 0.3 truecm
\end{flushright}

\begin{center}
{\bf More on strongly coupled quenched QED}
\vskip 0.6truecm
{\normalsize Maria--Paola Lombardo}\\
{\it
HLRZ c/o KFA J\"{u}lich, \\D-52425 J\"ulich, Germany}
\vskip 0.5truecm
{\normalsize Aleksandar Koci\'{c} and John ~B.~Kogut} \\
{\it Department of Physics,
University of Illinois at Urbana--Champaign,\\
1110 West Green Street, Urbana, IL 61801-3080, U.S.A.} \\
\vskip 0.5truecm
\today
\end{center}

\begin{abstract}
We study the critical region of lattice QED4
in the quenched approximation.   The issue
of triviality is addressed by contrasting
simulation results for $<\bar\psi\psi>$ and for
the susceptibilities
with the predictions of  two
critical scenarios -- powerlaw scaling,
and triviality \`a la Nambu--Jona Lasinio.
We discriminate among
the two possibilities with reasonable accuracy and
we confirm previous results for the critical point
and exponents thanks to new analysis
strategies and good quality data.
The interplay of chiral symmetry breaking with the
Goldstone mechanism
is studied in detail, and some puzzling features of past
results are clarified. Chiral symmetry restoration
is observed in the  spectrum:  the candidate
Goldstone boson decouples in the weak coupling phase,
while the propagators of the chiral doublets become
degenerate.
We  also present the first measurements of the full
mesonic spectrum, relevant for the study of flavour/
rotational symmetry restoration. The systematic effects associated
with our measurements are discussed in detail.
\end{abstract}
\end{titlepage}

\section{Introduction}

{}~~~~One of the most
fundamental and difficult questions in high energy theoretical
physics is whether theories which are strongly interacting at short distances
can exist. It is generally assumed that only asymptotically free theories
exist although the evidence for this point of view is based almost wholly
on perturbation theory. If a class of theories in which interactions are
strong at short distance does in fact exist, then new theoretical
approaches to symmetry
breaking in the Standard Model (the Higgs sector) would become feasible.
In addition, the
existence of Quantum Electrodynamics (QED) as a self-contained theory
would become a viable theoretical possibility and this fact would have
considerable impact on unification schemes ~\cite{INTRO}.

Past studies of lattice QED have
discovered a chiral symmetry breaking phase transition at a relatively large
value of the coupling constant ~\cite{CSB}.
This transition is second order in the
non-compact version of the theory ~\cite{DKK88}.
By combining accurate measurements of the
chiral condensate with spectroscopy calculations, an equation of state and
scaling laws were found
which are characteristic of an interacting, non-trivial underlying theory.

The results were particularly accurate for the quenched theory, due
to the simplicity of the underlying free dynamics~\cite{NMF91},
{}~\cite{KKLW93}.
Past work on quenched QED
studied the interplay of spectroscopy, scaling
laws and critical indices.
An equation of state, and scaling laws were derived
for the techni-meson masses by exploiting correlation length scaling.
The resulting universality relations were confirmed by simulations
and the anomalous dimension $\eta$ was measured to be approximatively
$0.50$ ~\cite{KKLW93} in good agreement with
past lattice simulations~\cite{NMF91} and hyperscaling relations, as
well as with the model's analytic solution ~\cite{EOS90}, ~\cite{BLM90}.

New analytic insights into chiral
transitions were also developed to make the most efficient and compelling use
of
the simulation data~\cite{KKL93}.

Despite these successes,
the consistency of the data with a logarithmically trivial equation of
state is  still an open problem. While corrections
a' la $\lambda \phi^4$
were found to be inconsistent with a fermionic theory
{}~\cite{KK93}, a
trivial theory based on a Nambu--Jona Lasinio model
\cite{NJL}  is  still possible.

In this work we address the issue of triviality in quenched QED by
discussing, and comparing in detail, the two possibile critical
behaviors -- power law scaling, and triviality a' la
Nambu--Jona--Lasinio.
We have performed new, extensive measurements for the chiral
condensate
and the susceptibility in the scalar channel. A detailed study of
finite size effects gave us good control over systematics errors.
The good quality of the data available for quenched QED,
together with old and new ``smart'' analysis strategies  enabled us
to discriminate between the two scenarios with reasonable accuracy.
Previous results for the critical point and exponents are confirmed,
and cross--checked by making use of the new numerical analysis
strategies, which complement the ones discussed in ~\cite{KKL93}.

Another central issue is the mechanism of chiral symmetry restoration
at the transition. The pion mass, measured in the past, displayed some rather
puzzling systematics: it decreased with $\beta$ even as we pass through
the chiral restoration transition. In the present work, we
study the behaviour of the scalar and pseudoscalar propagator in much
more
detail than in past work. The numerical results for the pion mass
are confirmed. Nevertheless, we observe clear indications of chiral symmetry
restoration, both in the scalar and in the
vector sector: the propagators of the chiral doublets become
degenerate,
and the amplitude of the Goldstone mode drops by an order of magnitude
across the transition. We speculate that in the thermodynamic limit,
the Goldstone mode would decouple completely at the chiral symmetry
restoration transition.

We have also obtained estimates for the sigma mass, which evaded us
in
previous measurements ~\cite{KKLW93}. Unfortunately, they are not accurate
enough
to be compared with the fermion mass, which would be relevant for the
triviality issue. We also present the first measurements of the full
mesonic spectrum, which should provide information on the restoration
of the rotational/flavor symmetry -- i.e. on the symmetries of the
continuum theory.

Finally, despite the simplicity of the quenched theory, the spectroscopy
computation, and the systematics of finite size effects contain several
pecularities. We collect these  issues in the
last section for the interested reader.

\section {Equation of state and scaling }

{}~~~~In our past work  we have discussed how the chiral equation of
state
can be derived, and used to obtain information about the nature of the
continumm limit defined at the chiral transition point.
Briefly, given that a direct computation of the renormalized charge
is plagued with both methodological and computational problems, one resorts
to the study of an effective action, which in turn provides information
on the critical exponents. The key point is the expression for the
renormalized charge~\cite{FFS}
\begin{equation}
g_R \simeq \xi^{(2\Delta - \gamma - d\nu)}
\end{equation}
where the critical indices in this expression should be familiar.
{}From the above expression, it is clear that a necessary condition
for $g_R$ to be different from zero in the continuum ($\xi \to
\infty$) limit is
\begin{equation}
2\Delta - \gamma - \nu = 0
\end{equation}
i.e. hyperscaling must hold.
Our strategy is therefore to identify the proper equation of state,
and compute the relevant exponents. (The obvious limitation is that
this implies a certain amount of guesswork, whose validity is not
always straightforward to verify a posteriori.)

In past work, the critical
exponents have been computed with good accuracy in the framework of
an equation of state derived in analogy with ferromagnetic
systems. The critical exponents are
different from mean field ones, and satisfy hyperscaling.
That hints, according to the previous discussion, at a non-trivial
continuum limit.

However in four dimensions, it is possible that
logarithmic corrections to scaling drive $g_R$ to zero. These
corrections would produce ''effective'' exponents distinct
from mean field, and they would still satisfy hyperscaling to good
accuracy. So the non--trivial behaviour observed when testing
the data against the hypothesis of power-law scaling could be
misleading.

This possibility was extensively studied in ~\cite{S91},
{}~\cite{GHLRSSW90}.
The authors however made use of an EOS based on a $\lambda \phi^4$ model,
ruled out  in ~\cite{KK93}, ~\cite{K94}
both on numerical and theoretical grounds.
We refer the reader to the original literature for discussions of
this point.

On the contrary, the theoretically motivated possibility of trivial
behavior realized \`a la Nambu--Jona Lasinio has not been studied in
detail.
Recall that
the continuum model must be the gauged Nambu--Jona Lasinio model, in
order for an analytic renormalization group to exist. That is, the theory
must be parametrized by two coupling constants, the electrodynamic and the
four fermi interaction strengths ~\cite{EOS90}, ~\cite{BLM90}.
{}From this point of view, the question is if the gauge fields are
enough to `promote' the trivial Nambu--Jona Lasinio model to an interacting
model, characterized by power law scaling with non mean field
exponents,
or if, instead, triviality survives the introduction of the gauge interaction.

To our knowledge,
the only work which addresses the possibility of trivial logarithms
\`a la Nambu--Jona Lasinio in QED  is ~\cite{H90}, where this
possibility
was first put forward, but the numerical study  made use of data
obtained with very limited
resources (small lattices) and was not compelling.

We thus decided to explore this issue in detail, and in the
following we will consider in parallel
the power--law equation of state
and the NJL log-corrected mean field.

 To make
this short exposition self contained, we recall that the first one
can be derived by the chiral Equation of State written in a standard
form, under the assumption of scaling
\begin{equation}
m_e = <\bar\psi\psi>^\delta f(t/<\bar\psi\psi>^{1/\beta_{mag}})
\end{equation}
Its first order approximation reads
\begin{equation}
m_e = a^P (\beta_c - \beta) <\bar\psi\psi>^{\delta - 1/ \beta}+
b^P <\bar\psi\psi>^\delta
\end{equation}
Past work has shown that $\delta - 1/\beta$ (the exponent $\gamma$)
is, with excellent accuracy, 1. The characteristics of the universal functions
$f$ associated with such a value of the $\gamma$ exponent were discussed in
\cite{KKL93}, and verified in \cite{KKLW93}. Our candidate equation of state
for a possible non-trivial cutoff-free theory will thus be
\begin{equation}
m_e = a^P (\beta_c - \beta) <\bar\psi\psi>+b^P <\bar\psi\psi>^\delta
\end{equation}

Coming to the log-corrected mean field behaviour, we study the NJL EOS
\begin{equation}
m_e = a^{NJL} (\beta_c - \beta) < \bar\psi\psi> +
b^{NJL}<\bar\psi\psi>^3log<\bar\psi\psi>
\end{equation}
which can be derived from the large-N expansion of a four-fermi lagrangean
{}~\cite{H90}, ~\cite{KKK94}.

Once more, we stress that it is important to notice the
position of the logarithms, which dictate the sign of the scaling violations:
as discussed in~\cite{KK93},~\cite{KKK94}   their positions are
independent of approximations and
are, in fact, generic to any fermionic model.

We will comment at the end of this Section on the possible corrections
to this leading-log behaviour.

\subsection {Equation of state fits}

{}~~~~As discussed above, we have compared our data
with the two EOS's

\begin{equation}
m_e =  a^P(\beta_c - \beta)<\bar\psi\psi> + b^P<\bar\psi\psi>^\delta
\end{equation}

and
\begin{equation}
m_e =  a^{NJL}(\beta_c - \beta)<\bar\psi\psi> +
b^{NJL} <\bar\psi\psi>^3log<\bar\psi\psi>
\end{equation}

We have used our old data for the chiral condensate
obtained on the $24^4$ lattice, for $\beta$
ranging from $.260$ to $.235$, $\Delta \beta = .001$, and five bare
electron mass values
equi--spaced between $.001$ and $.005$.
 We have fitted the data for several
$\beta$ ranges. The inclusion of $\beta < .240$ spoils the power-law
fits. Apparently such strong couplings are too far from the critical
point to be described by a simplified EOS.
The fits, however, are satisfactory and stable
once $\beta$ is restricted to the range $.240$ to $.260$.
We show in Fig.~1
the results of the fit for the power law form for two $\beta$
intervals
:[.250,.260], and [.240,.260].
Consider, for instance, the fit over the
interval [.250 .260], which includes only 10 $\beta$ values. Its validity
extends well beyond the fitted interval, and its results agree with
the ones obtained by enlaring the $\beta$ interval.

The fit \`a la Nambu--Jona-Lasinio (Fig.~2), on the other hand,
are qualitatively poorer, and the resulting parameters are sensitive to
the $\beta$ interval chosen.
These results are summarized in the Table ~\ref{FIT}.

 The coefficient
of the term linear in $<\bar\psi\psi>$  is
the same for the two models.
Note that for both models the coefficient $a$ can be expressed as
\begin{equation}
a = 1/ <\bar\psi\psi> (\partial <\bar\psi\psi> / \partial \beta)
(\partial m_e / \partial <\bar\psi\psi> )
\end{equation}
i.e. it can be shown explicitly that $a$ is independent of
the nature of the critical singularities,
and it could be read off the data in
an `effective analysis' style, for either EOS.
Differences in the two parametrizations come from the chiral
condensate's behaviour
close to the critical point.

The fits clearly favour power--law behavior, with
$\beta_c  = .2573(1)$, $\delta =  2.13(1)$, in agreement with previous
findings (the errors have been estimated by jack-knifing the
results obtained, discarding one point at a time).

\begin{table}
\caption{Results of the EOS fits}
 \begin {tabular} {|l|l l l l l l l|} \hline
$\beta$ range  & $a^{NJL}$&$b^{NJL}$&$\beta_c^{NJL}$&
 $a^{P}$&$b^{P}$&$\beta_c^{P}$&$\delta$\\
  \hline
.250 -- .260&-5.43(4)&-2.76(4)&.2542(1)&-5.40(2)&1.04(14)&.2572(6)&2.12(8)\\
.245 -- .260&-5.41(3)&-2.59(3)&.2538(1)&-5.42(2)&1.05(2) &.2572(1)&2.13(1)\\
.240 -- .260& --     & --     & --     &-5.54(2)&1.07(2)&.2573(1)&2.13(1)\\
\hline
 \end{tabular}
\label{FIT}
\end{table}

\subsection{The exponent triangle}

{}~~~~We show here that the results for the
critical exponents can be obtained very nicely in  graphic form,
under the {\it single} assumption of the existence of an Equation of State
which obeys scaling:
\begin{equation}
t =  <\bar\psi\psi> ^{1/\beta_{mag}} F(m/<\bar\psi\psi>^\delta)
\end{equation}
or, equivalently:
\begin{equation}
m =  <\bar\psi\psi> ^{\delta} f(t/<\bar\psi\psi>^{1/\beta_{mag}})
\end{equation}
It is convenient to notice  that these two forms are related by the
`symmetry' transformation:
\\
\noindent
$m \longleftrightarrow t$ \\
$\beta_{mag} \longleftrightarrow 1/\delta$\\
$f \longleftrightarrow F$\\
\\
\noindent
(a simple example of this transformation is of course
\\
$<\bar\psi\psi>_{t=0} \propto  m^{1/\delta}
\longleftrightarrow
<\bar\psi\psi>_{m=0} \propto  t^{\beta_{mag}}$.)

\noindent
Consider now the two logarithmic derivatives
\begin{equation}
R_t = \frac{t}{ <\bar\psi\psi>}
\frac{\partial<\bar\psi\psi>}{\partial t},
\,\,\,
R_m = \frac{m}{<\bar\psi\psi>}\frac{\partial<\bar\psi\psi>}{\partial m}
\end{equation}

$R_t$ can be computed from the EOS, eq. (10)
\begin{equation}
1/R_t = 1/\beta_{mag} - \delta y F'(y) / F(y)
\end{equation}
where $y  = m /<\bar\psi\psi>^\delta.$
The 'symmetric' expression for $R_m$ reads:
\begin{equation}
1/R_m = \delta - 1/\beta_{mag} x f'(x) / f(x)
\end{equation}
where $x  = t /<\bar\psi\psi>^{1/\beta_{mag}}$.

$R_m$ can also be expressed as a function of the susceptibilities
\begin{equation}
\chi_{\sigma} = \int_x <\bar\psi\psi(x) \bar\psi\psi(0)>,\,\,\,
\chi_{\pi} = \int_x <\bar\psi i \gamma_5 \psi(x) \bar\psi i \gamma_5\psi(0)>
\end{equation}
which are the zero momentum projections of the scalar and pseudoscalar
propagators. These two susceptibilities are related to the order
parameter via : $\chi_{\sigma} = \partial <\bar\psi\psi>/ \partial m$
and the Ward identity $\chi_\pi = <\bar\psi\psi>/m$. These
relationships
were discussed in ~\cite{KKL93}, and we will exploit them in the next Section.
Here,
we will simply make use of the numerical derivatives of the order parameter.

Using Eq.(10) (11) and the symmetry transformation,
it is easy to verify that $R_t$ and $R_m$ satisfy:
\begin{equation}
\frac {R_t}{\beta_{mag}}  + \delta R_m = 1
\end{equation}

We emphasize that this result depends only on the scaling
form of the EOS.

When $R_m$ is 0, in the chiral limit of the strong coupling phase,
$R_t$ defines the exponent $\beta_{mag}$. The critical point corresponds
to $R_t = 0$, and correspondingly we have $R_m = 1/\delta$. Finally,
$R_t$ is $-\gamma (= \beta_{mag}(\delta -1))$
in the chiral limit of the weak coupling phase
$(R_m = 1)$.

Thus,
it is possible to read off
the exponents $\beta_{mag}$, $\delta$ and $\gamma$ from the plot $R_t$ vs.
$R_m$ which itself only requires $\beta_c$ as input.

We show the 'triangle plot' in Fig.~3 for the $24^4$ lattice.
On this lattice we have a grid of data fine enough to permit the accurate
calculation of numerical derivatives of the chiral condensate. Since only the
global averages and uncertainties were saved,
we can not make comprehensive error estimates
for the derivatives, so we include only the errors on the chiral
condensate itself in the plot.
Since the results for different values of the parameters
are strongly correlated (remember that the same gauge configurations
were used for all the parameter values), we are confident that the
errors on the derivatives are small.

We see in the plot (Fig.~3) that the data points lie on a straight line,
in very good agreement with the
fit prediction, and the exponents can be read off the axes,
as anticipated.

\subsection{Susceptibilities analysis}

{}~~~~In ref. ~\cite{KKL93} we have discussed the
properties of the ratio $R_m  (t,m)=
m/ <\bar\psi\psi>\times \partial <\bar\psi\psi> / \partial m$ .
It was shown how its behaviour is dictated by symmetry
arguments, and provides information on the critical point, and
critical exponents, with no a priori assumptions. In particular,
$R_m (t,0) = 0(1) $ in the strong (weak) coupling limit and
$R_m(0,m) = 1/\delta$.
For power law scaling $R_m$ is thus expected to be constant
with $m$ and equal to $1/\delta$ at the critical point, while for the
NJL model it would
follow the `effective $\delta$, $R_m (0,m) = 1/\delta_{eff} = 1/ (3 +
1/log<\bar\psi\psi>)$. $R_m$ can thus be used to discriminate among
different critical behaviours.

Motivated by these considerations, in this new round of
simulations we have measured the scalar susceptibility by making
use of a noisy estimator. The details of these measurements can be
found in the last Section. As we shall show,
the new data for the scalar susceptibility provides an independent check
on the results obtained above.

We show in Fig.~4a the ratio of susceptibilities plotted
at fixed $\beta$ as a function of the bare electron mass. We plot
our results on the $16^4$ lattice, where the sigma susceptibility
was ``directly'' computed and, as a cross check, the results
on the $24^4$ lattice, where the sigma susceptibility was obtained
by numerical differentiation, wherever they overlap.
Note the nice
agreement in the critical region (see also the zoomed
view Fig.~4b where we have included all the available
$\beta$'s on the big lattice), and the (small) discrepancy at strong
coupling which we will discuss at the end. This discrepancy is of
course
irrelevant for the issue of the critical behaviour.
The straight line is
drawn giving $1/\delta$ as estimated by the fit, and falls, as
it should, half way between $\beta = .255$ and $\beta = .260$. In Fig.~5
 we test the susceptibility data against the prediction of the
Equation of State \`a la Nambu--Jona--Lasinio. To do so, it is more
convinient to plot $1/R_m - 3$ versus $1/<\bar\psi\psi>$ since this
is predicted to be a straight line with unit slope, which is drawn
as a solid line. Clearly, the data does not follow it. (It could be
that the inclusion of a scale in the log changes the trend in the
right direction. However, we have tried to include a scale in the
direct fits to the equation of state discussed above, and the results
do not change qualitatively. In particular, the scale and the
parameter $b$ are sensitive to the width of the $\beta$ interval.)
The dashed line, again, is the power law prediction
:
we learn that the susceptibility data favors powerlaw scaling.

\subsection{EOS's higher order corrections}

{}~~~~Finally, a word of criticism is in order.
The above results have been obtained by assuming
that the theory at the critical point is well described by an effective
model written in terms of the chiral condensate. Besides that,
we restricted ourselves to the simplest parametrizations of the
two universality classes we have considered. It could be
that the difference in behaviour we have observed between power law scaling
and log-corrected mean field are due to an inadequate parametrization of the
trivial EOS.
In particular, the derivation of the $NJL$ EOS
suggests that the effective theory should be
formulated in terms of the renormalized electron mass, as
opposed to the chiral condensate. Unfortunately, the quality of the fermion
mass data is not adequate for an EOS study , but we can try to get
a feeling of the magnitude of the corrections.
We expect the dependence of the chiral condensate on the dynamical
fermion mass is `somewhere  between' the strong coupling limit and
pure free field behavior (we stress again that we are only making
an `order of magnitude' estimate).

The leading term of the strong coupling expansion ~\cite{KL94}
lagrangean  is given by
\begin{equation}
S_{QED S.C} = 1/4 (\bar\psi\psi)_x (\bar\psi\psi)_{x+\mu}
\end{equation}

which gives,

\begin{equation}
2<\bar\psi\psi> + m_e = M_F
\end{equation}

The analogous relationship in the free field limit was discussed
in ~\cite{S91} and reads:

\begin{equation}
<\bar\psi\psi> = .62M_f
\end{equation}

We then plot (Fig.~6) the fermion mass versus the chiral condensate.
We see that the numerical results lie between the two
limiting behaviors
(we can ignore $m_e$ in the plot at strong coupling),
and that, in our region
of masses, the corrections induced by the replacement of the
chiral condensate by the dynamical fermion mass in the
Equation of State are less than the statistical errors.

\section{Spectroscopy }

{}~~~~We discuss the results obtained on a $16^3 \times 32$ lattice, at
$\beta$ = (.250 .255 .260 .270 .280),
for bare quark masses ranging from
$.003$ to $.025$. For a very limited subsample of parameters we also took
data on a $16^3\times 64$ lattice, for checking purposes.
We have used a variety of sources, which allows us to gain better
control and understanding of our results. We have used a standard
point source and a wall-noisy source. We have also
implemented the sources for the measurements of the full mesonic
spectroscopy, which we discuss in the next Section. This
allows us to cross--check the results for the
pion, which couples to all the sources we
have used, and to understand the systematics affecting each
measurement technique. For the $\sigma$ particle,
the use of a point/wall-noisy source was mandatory, since as we
have discussed in \cite{KKLW93} this particle does not couple
to a rigid wall in the quenched case. Here we were able to
obtain reasonable sigma propagators, even if the extraction
of the sigma mass was problematic, as we discuss below. We will give
the details of the analysis in the last Section, where we will also
collect all the Tables.

\subsection {Chiral symmetry restoration and
level ordering}

{}~~~~The analysis of the scalar and pseudoscalar propagators provides
information on the mechanism of chiral symmetry breaking and the
appearance of a Goldstone mode in the meson spectrum.

The main qualitative feature we observe in the (pseudo)scalar
sector of the spectrum is the following: the mass of the lowest
pseudoscalar state decreases with $\beta$ at fixed
quark mass, while the mass of the first excited state in the
pseudoscalar propagator is always comparable
with the sigma mass. Also, in past work we have studied the property
of the pion mass, and found that the deviations from Goldstone
behaviour
were not dramatic in the weak coupling phase.

At first sight, this
seems puzzling and at odds with chiral symmetry restoration at the
transition, where one expects that the sigma and the pion masses become
degenerate, and that the pion is no longer
a Goldstone particle. One possible solution of this apparent paradox, and a key
to gaining better understanding of the mechanism of chiral symmetry
restoration,
relies on a more detailed analysis of the propagator's behaviour. As we now
show, in
the weak coupling phase the amplitude of the excited states in the
pseudoscalar channel becomes dominant, while the fundamental one
is suppressed. This explains how the low-lying
state in the pseudoscalar channel maintains the property
of a Goldstone boson beyond the transition point, as noticed in
past work, but there is no contradiction with the Goldstone Theorem.
The candidate Goldstone boson simply decouples in the weak
coupling phase.

The Tables ~\ref{taba1},~\ref{taba2} , and the figures 7-8,
show the  amplitudes of the fundamental and
first excited states. In the tables the propagators were normalized
to $G_\pi(t=0)$, so the quoted amplitudes represent the fraction that
each state (Goldstone, 1st excited) contributes to the complete
propagator. The Goldstone contribution $A_G$ is about 70 \%
at $ m = .003$ and $\beta = .250$, while it is only 10 \%
at the same mass value, and $\beta = .280$. Around the critical point
the Goldstone amplitude become dominant, in the small ( $\le .01$)
mass limit, and does not depend much on the mass itself. For higher masses,
the two amplitudes are rather insensitive to $\beta$, as expected.
Note that 1 is an upper bound for  $A_G + A_{1st excited} $ ,
because of positivity, while $1 - (A_G + A_{1st excited}) $ represents
the contribution of higher excitations, which is small.
The amplitude itself (i.e., without normalization)
$G_\pi(t=0)$  is a rather smooth function of
$\beta$ and mass. For instance,
at $\beta = .250 (.280)$ it goes
from 3.61(1.71) at $m = .003$ to 2.61(1.68) at $m = .009$.
In Fig.~7
we plot the amplitude in the pion channel, and the total amplitude minus
the amplitude in the pion channel, for $m = .003$ and $m = .009$.
The change in the trend around the transition is very clear, the effect
being more pronounced at small masses, as it should be.

 Apparently,  in the chiral symmetric phase
the Goldstone mode simply disappears, and the mass of the pseudoscalar
meson (the first excited state) becomes degenerate with the
sigma mass, consistent with chiral symmetry restoration. This
is certainly what ultimately happens in the perturbative limit.
The amplitude of the Goldstone channel apparently is not related
to the lattice geometry (at least, it does not change in a significant
way on the $16^3\times 64$ lattice), but it might be controlled by
the lattice size. It is very possible that on a larger lattice
the suppression of the Goldstone amplitude is abrupt. It would be
interesting to check this point.

Another point to be considered  is that on the lattice chiral symmetry should
be realized at the level of propagators  -- i.e. the sigma and pion
propagators are related by a lattice chiral symmetry transformation.
{}From this point of view, the mass degeneracy is a by-product of
a more general symmetry, and it is not suprising that the correct
behaviour at long distance is the last one to be recovered.
To study the short-distance behaviour of the restoration of chiral
symmetry, we can consider the zero distance propagator.
For instance, we show in Fig.~8  the ratio of the zero distance propagators
$G(0)$ in the scalar and pseudoscalar
channel plotted versus $\beta$ at fixed electron mass:
the behaviour is the one expected on the
grounds of chiral symmetry, and it is analogous
to the trend of the susceptibility ratio.

In a completely analogous way we can study the (pseudo)vector sector. The
pseudovector propagators are rather intractable, and we
do not have a good estimate for the A1 mass. But again,
in order to study chiral symmetry breaking/restoration,
we can use, as in the scalar/pseudoscalar case, the propagators themselves.
Fig.~9 shows the
results for the ratio of the $\rho$ to the $A1$ propagators
at zero distance. Again, chiral symmetry restoration is quite visible
for $\beta > .260$.

Another important issue is the relationship between the $\sigma$ mass
and the fermion. From the data we quote (Table~\ref{sigm},
Table~\ref{femt})
the $\sigma$ mass is definitively
heavier than twice the fermion mass. Naive Nambu-Jona-Lasinio behavior
predicts $\sigma = 2 m_f$, while $\sigma < 2m_f$
could be interpreted as a signature of non--triviality (there is nonzero
binding enery in the sigma channel). Our simulation results do not fit
into either scenario. However, our estimate of the $\sigma$ mass
is effected by the uncertainties discussed in the last Section:
basically, the long distance behaviour was not clear enough, and our
results
come from the analysis of intermediate times. The ordering $m_\sigma >
2m_f$
suggested by our data should not be taken as conclusive.

\subsection {Full mesonic spectrum and the restoration of the
flavor/rotational symmetry}

{}~~~~All the symmetries of the continuum have to be realized at the chiral
transition, if there is a bona fide continuum limit at that point.
Golterman ~\cite{G86} has constructed the rest-frame meson operators,
classified them according to the lattice symmetry group, and found
relationships with the  continuum operators.
We borrowed these operators for our QED computations. They have been
used first in QCD by the High Energy Monte Carlo Grand Challenge group.
We refer the reader to the original works for more detail,
and just recall here that the Dirac matrix $\gamma$ and $\xi$
act in spinor and flavor space, respectively. In this notation the
Goldstone meson is excited by the operator
$\gamma_4 \gamma_5 \xi_4 \xi_5$.

We have analyzed the data for these operators along with the spectrum data
discussed above, on the same $16^3\times 32$ lattices.
The extraction of the direct component of the
propagators was necessary to obtain a
satisfactory plateau for the effective masses.

All the (pseudo)vector particles turned out to be degenerate for the
full set of parameters we have explored, and we will not comment
further on them.

The scalar sector is more interesting.
We display a representative sample of plots for the effective masses
in
Fig.~10. The curves level off for relatively small t.
We chose t=5 as a safe starting point for the weighted average
of the results.
The quality of the data for the pseudoscalar mesons
is operator dependent as can be seen from the magnitude of
the errors in the Tables~\ref{tabm03}, ~\ref{tabm09}.
We show only a subset of the results, to illustrate their essential features.

Basically, all
the non-Goldstone mesons turn out to be mutually (almost) degenerate.
A certain trend towards the restoration of continuum symmetry can
be observed, especially for the meson excited by $\gamma_5\xi_k\xi_5$,
the most significant for the study of flavor symmetry restoration.
This trend is shown in Fig.~11.
However, it is not clear if this trend toward degeneracy, apparent
from the data and from the figures, is induced by the chiral transition, or
by the perturbative limit.

\section{Numerical details, discussion of finite size effects and
of the spectroscopy analysis}

{}~~~~As anticipated, this Section is devoted to the discussion of
the details of the new simulations, and contains all the
relevant Tables.
In the new simulations
we have used the same algorithm as in the past (see \cite {ALGO}).
It begins in momentum space, and produces the appropriate Gaussian
distribution of photons. Then, using a Fast Fourier Transform it generates
a set of dimensionless gauge fields in coordinate space. The coupling
of the gauge fields to the electrons is then implemented by an appropriate
rescaling of the gauge fields.

\subsection{Chiral condensate and sigma susceptibilities}

{}~~~~The data for the chiral condensate used in this paper are from our old
simulation on the $24^4$ lattice, and from new simulations on the
$16^3\times 32$, $16^4$ and $8^4$ lattices.

 The chiral condensate was measured
by inverting the Dirac operator with a noisy source defined on the
even sites of the lattice. We have also
measured  $\frac {\partial <\bar\psi\psi>}{\partial m}$ (the scalar
susceptibility, $\chi_\sigma$), by using the same noisy background as for
the chiral condensate. Several Tables
(~\ref{chip8}--~\ref{chis32}) collect our results.

As is well known, the finite size effects on the order
parameter (and consequently
its derivative, the sigma susceptibility) are not dramatic. However, they
are not negligible, and have some peculiar characteristics
which should be discussed. First,
the qualitative behaviour of
the  finite size/geometry effects depend on the phase--whether chiral
symmetry is broken or restored.

\begin{table}
\caption {Chiral condensate on the $8^4$ lattice}

 \begin {tabular} {|l|l l l l l l l|} \hline
 & .250 & .255 & .260 & .265 & .270 & .275 & .280 \\
  \hline
 .003&.1036(64)&.0786(53)&.0630(38)&.0453(27)&.0349(25)&.0294(15)&.0236(10)
 \\
 .007&.1331(52)&.1132(45)&.0989(37)&.0781(26)&.0645(20)&.0585(20)&.0512(20)
 \\
 .011&.1501(41)&.1338(33)&.1209(35)&.1034(27)&.0883(20)&.0805(19)&.0710(17)
 \\
 .015&.1606(30)&.1473(30)&.1340(26)&.1201(26)&.1094(21)&.0989(20)&.0887(17)
 \\
 .019&.1792(30)&.1665(28)&.1511(26)&.1395(25)&.1267(21)&.1172(20)&.1042(17)
 \\
 .023&.1915(30)&.1779(27)&.1673(26)&.1508(23)&.1360(19)&.1303(21)&.1173(18)
 \\
 .027&.2031(27)&.1866(25)&.1730(23)&.1614(22)&.1543(19)&.1440(21)&.1316(17)
 \\
 .031&.2121(28)&.1984(27)&.1891(26)&.1722(21)&.1609(19)&.1542(19)&.1417(18)
 \\
 .035&.2238(27)&.2074(24)&.1958(23)&.1837(22)&.1709(21)&.1634(20)&.1552(21)
 \\
 .039&.2256(27)&.2171(25)&.2064(27)&.1939(23)&.1806(19)&.1751(20)&.1634(18)
 \\
 .043&.2351(25)&.2205(23)&.2100(22)&.1997(21)&.1931(20)&.1814(20)&.1714(18)
 \\
 .047&.2426(23)&.2343(23)&.2209(21)&.2113(22)&.2023(21)&.1916(19)&.1803(17)
 \\
 .051&.2495(24)&.2403(23)&.2299(23)&.2168(20)&.2056(20)&.1981(20)&.1877(18)
 \\
 .055&.2592(23)&.2447(23)&.2346(21)&.2239(20)&.2164(18)&.2079(19)&.1974(17)
 \\
 \hline
 \end{tabular}

\label{chip8}
\end{table}

\begin{table}
\caption {$m_e \times \chi_{\sigma}$ on the $8^4$ lattice  }
 \begin {tabular} {|l|l l l l l l l|} \hline
 & .250 & .255 & .260 & .265 & .270 & .275 & .280 \\
  \hline
 .003&.0359(43)&.0381(31)&.0371(21)&.0324(13)&.0259(12)&.0244( 8)&.0212( 6)
 \\
 .007&.0364(37)&.0417(33)&.0454(22)&.0483(14)&.0457(12)&.0435(14)&.0410( 9)
 \\
 .011&.0421(29)&.0507(22)&.0530(21)&.0564(15)&.0565(13)&.0572(11)&.0537(10)
 \\
 .015&.0586(19)&.0562(21)&.0599(16)&.0630(16)&.0655(11)&.0645(12)&.0623( 9)
 \\
 .019&.0594(22)&.0632(19)&.0656(17)&.0676(15)&.0710(13)&.0701(11)&.0702(11)
 \\
 .023&.0629(19)&.0683(16)&.0686(18)&.0723(14)&.0759(11)&.0742(11)&.0757(10)
 \\
 .027&.0688(20)&.0726(17)&.0769(15)&.0779(14)&.0777(13)&.0802(13)&.0812(11)
 \\
 .031&.0692(19)&.0742(18)&.0772(17)&.0809(14)&.0837(12)&.0837(12)&.0861(10)
 \\
 .035&.0760(20)&.0780(17)&.0806(16)&.0852(15)&.0871(13)&.0871(12)&.0884(11)
 \\
 .039&.0779(18)&.0819(16)&.0840(16)&.0877(14)&.0887(12)&.0917(12)&.0917(11)
 \\
 .043&.0822(16)&.0831(16)&.0862(15)&.0902(14)&.0920(12)&.0927(12)&.0954(11)
 \\
 .047&.0831(16)&.0864(16)&.0888(14)&.0906(14)&.0950(13)&.0946(12)&.0985(11)
 \\
 .051&.0852(16)&.0884(16)&.0894(16)&.0941(13)&.0981(12)&.0973(13)&.1003(11)
 \\
 .055&.0870(16)&.0905(16)&.0955(13)&.0978(13)&.0972(13)&.0997(14)&.1034(12)
 \\
 \hline
 \end{tabular}

\label{chis8}
\end{table}

We discuss first the behaviour in the strong coupling phase.
We plot in Fig.~12 the data for the chiral condensate
on the three symmetric lattices. The results on the small lattice are
effected by finite size effects of the order of several percent, but
on all the lattices the chiral condensate behaves in a qualitative
correct way.  The only perplexing point is that the chiral condensate
increases with decreasing volume. One should expect the opposite trend,
since chiral symmetry breaking, which produces a non-zero value of the
chiral condensate in the chiral limit, is an infinite volume effect which
should be obscured on a finite lattice.  However, from
Fig.12 we observe that the derivative of the chiral
condensate with respect to the quark mass  is increasing as the
volume decreases at smaller masses.  This effect can be read off
also from the susceptibility data.
(This is consistent with the `early' transition observed for small
volume: finite volume increases the slope of the chiral condensate
as a function of the bare quark mass, thus mimicking the approach to
the critical point.)
If this trend is maintained,
the curves for the chiral condensate
are going to `cross' at some point, thus recovering the expected pattern
for finite size effects in the chiral limit-- that the chiral condensate
decreases with
volume. Indeed,
qualitative arguments leading to the conclusion that the chiral
condensate should decrease with volume,  require that the dominant contribution
to the chiral condensate is from spontaneous (as opposed to the explicit)
symmetry breaking.
On small lattices, and large masses, the main contributions
to the chiral condensate/susceptibility come from the excited states, so
the standard arguments do not necessarily apply. Another possibility is that
finite volume effects push up the physical pion mass, in such a way
that the pion on a $8^4$ lattice is even heavier than the one on a $16^4$.
Then PCAC would tell us that the chiral condensate increases with volume,
if $f_\pi$ is (almost) constant with volume.

\begin{table}
\caption {Chiral condensate on the $16^4$ lattice}
 \begin {tabular} {|l|l l l l l l l|} \hline
 & .250 & .255 & .260 & .265 & .270 & .275 & .280 \\
  \hline
 .003&.0977(21)&.0711(17)&.0558(14)&.0412(10)&.0316( 6)&.0258( 4)&.0214( 3)
 \\
 .007&.1189(17)&.1016(13)&.0866(12)&.0725(10)&.0620( 8)&.0529( 7)&.0463( 5)
 \\
 .011&.1407(14)&.1250(12)&.1091(10)&.0971( 9)&.0827( 7)&.0745( 7)&.0663( 6)
 \\
 .015&.1606(12)&.1418(11)&.1286(11)&.1146(10)&.1032( 9)&.0929( 7)&.0842( 6)
 \\
 .019&.1726(14)&.1577(12)&.1430(10)&.1313(10)&.1192( 8)&.1097( 7)&.0990( 6)
 \\
 .023&.1843(12)&.1687(11)&.1555( 9)&.1453(10)&.1325( 8)&.1236( 8)&.1131( 7)
 \\
 .027&.1955(11)&.1833(11)&.1688(10)&.1554( 9)&.1448( 8)&.1351( 8)&.1266( 7)
 \\
 \hline
 \end{tabular}
\label{chip16}
\end{table}

\begin{table}
\caption {$m_e \times \chi_{\sigma}$ on the $16^4$ lattice  }
 \begin {tabular} {|l|l l l l l l l|} \hline
 & .250 & .255 & .260 & .265 & .270 & .275 & .280 \\
  \hline
 .003&.0238(16)&.0304(10)&.0297( 9)&.0291( 5)&.0266( 3)&.0233( 2)&.0203( 2)
 \\
 .007&.0414(11)&.0434( 8)&.0445( 7)&.0455( 5)&.0440( 4)&.0419( 3)&.0391( 3)
 \\
 .011&.0492( 9)&.0527( 8)&.0546( 7)&.0563( 5)&.0552( 4)&.0541( 3)&.0520( 3)
 \\
 .015&.0564( 9)&.0603( 7)&.0621( 6)&.0640( 5)&.0634( 5)&.0627( 5)&.0617( 3)
 \\
 .019&.0633( 8)&.0664( 7)&.0694( 6)&.0706( 6)&.0710( 5)&.0703( 4)&.0695( 4)
 \\
 .023&.0685( 8)&.0714( 7)&.0745( 6)&.0761( 6)&.0765( 5)&.0774( 5)&.0760( 4)
 \\
 .027&.0735( 7)&.0756( 7)&.0782( 6)&.0806( 5)&.0814( 5)&.0818( 5)&.0817( 4)
 \\
 \hline
 \end{tabular}

\label{chis16}
\end{table}

\begin{table}
\caption {Chiral condensate on the $16^3 \times 32$ lattice}
 \begin {tabular} {|l|l l l l l |} \hline
 & .250 & .255 & .260 & .270 & .280 \\
\hline
 .003&.0980(18)&.0783(26)&.0633(14)&.0435( 9)&.0312(10)
 \\
 .005&.1116(17)& -- & -- &.0560( 9)&.0418( 8)
 \\
 .007&.1244(16)& -- & -- &.0664( 9)&.0524( 9)
 \\
 .009&.1323(15)&.1143(18)&.1002(12)&.0795(10)&.0606( 8)
 \\
 .011& -- & -- & -- &.0895(20)& --
 \\
 .013& -- & -- & -- &.0959(16)& --
 \\
 .015& -- &.1446(21)&.1280(12)&.1046(18)& --
 \\
 .017& -- & -- & -- &.1141(17)& --
 \\
 .019&.1739(28)& -- & -- &.1237(18)& --
 \\
 .021&.1774(22)&.1666(18)&.1507(11)&.1276(18)& --
 \\
 .023&.1840(25)& -- & -- &.1337(15)& --
 \\
 .025&.1904(21)& -- & -- &.1407(17)& --
 \\
 \hline
 \end{tabular}

\label{chip32}
\end{table}

\begin{table}
\caption {$m_e \times \chi_{\sigma}$ on the $16^3 \times 32 $ lattice  }
 \begin {tabular} {|l|l l l l l |} \hline
 & .250 & .255 & .260 & .270 & .280 \\
  \hline
 .003&.0237( 9)&.0219(13)&.0235( 7)&.0205( 5)&.0167( 6)
 \\
 .005&.0324( 8)& -- & -- &.0302( 5)&.0254( 6)
 \\
 .007&.0375( 9)& -- & -- &.0381( 5)&.0339( 5)
 \\
 .009&.0435( 8)&.0465(10)&.0477( 6)&.0452( 4)&.0409( 5)
 \\
 .011& -- & -- & -- &.0503( 8)& --
 \\
 .013& -- & -- & -- &.0575( 8)& --
 \\
 .015& -- &.0599(10)&.0615( 6)&.0612( 8)& --
 \\
 .017& -- & -- & -- &.0650(10)& --
 \\
 .019&.0614(13)& -- & -- &.0688( 9)& --
 \\
 .021&.0652(12)&.0704( 9)&.0714( 5)&.0710( 8)& --
 \\
 .023&.0671(12)& -- & -- &.0750( 7)& --
 \\
 .025&.0708(11)& -- & -- &.0785( 9)& --
 \\
 \hline
 \end{tabular}

\label{chis32}
\end{table}

So, we can find several qualitative explanations for this seemingly
puzzling behaviour. However, it should be noticed that in the unquenched
model,  the behaviour is the
conventional one, so it is possible that these results point out
some characteristic/pathology of the quenched model not yet explored.
The sigma susceptibility, again in the broken phase,
is very sensitive to finite volume effects: at $\beta$ = .250, mass = .003,
the difference between the results on the $8^4$ and on the $16^4$ lattice is
around $30 \%$,
as opposed to the $10 \%$
difference for
the chiral condensate. This can be understood by recalling that
in the strong coupling, infinite volume limit, the sigma susceptibility
is 0 for every value of the electron mass, while it is necessarily finite
on a finite volume. So, roughly speaking, we can expect
$\chi_\sigma \simeq 1/V$, which justifies the considerable sensitivity of the
sigma susceptibility to the volume.

The geometry effects, as deduced from the comparison of the results
on the $16^4$ and $16^3 \times 32$ lattice are instead very small,
thus showing that the long-distance behaviour in the strong coupling
phase does not have any peculiarity.

Consider now the weak coupling phase.
The pattern of finite
size effects in the chiral condensate is analogous to the one
observed at strong coupling, so apparently the finite size effects do
not 'see' the transition. The finite size effects
on the sigma susceptibility are instead different, and not as strong
in the weak coupling phase. Indeed, here the chiral condensate
should extrapolate to 0, and its derivative is supposed to have a
weaker
volume dependence in this phase.

In fact, finite volume corrections can be written as $<\bar\psi\psi> =
<\bar\psi\psi>(V=\infty, m_e)(1 - A/V^\alpha)$. When $<
\bar\psi\psi>=
m_e^x$, finite volume effects on its derivative at finite $m_e$ are still
$O(1-A/V^\alpha)$. The finite volume effects on the chiral
condensate and on its derivative are comparable, and thus smaller, for
the derivative,  than the ones
observed in the strong coupling phase.

The most peculiar systematics we have observed in the weak coupling phase
come from the small mass results  on the  asymmetric lattice (Fig.~13).
They look
at odds with chiral symmetry restoration, which is instead apparent
even on the smaller lattice.
We attribute them to a lack of complete suppression
of the Goldstone mode in the weak coupling phase, as we have
discussed above.

Summarizing, the finite volume effects, as studied on symmetric lattices,
are most important for the sigma susceptibilities in the strong
coupling phase, and can be traced back to the difficulties of a finite
lattice breaking chiral symmetry. They are not very strong,
but somehow puzzling, for the chiral condensate itself.

The geometry effects
are instead most significant in the weak coupling phase, at small masses,
thus demonstrating a rather complicated, and maybe not completely
understood, long distance behaviour.

Anyway, the data for the chiral condensate obtained on the $16^4$ lattice
are reasonably free of systematics in the entire range of
parameters we have explored, as can be inferred from their comparison
with
the $24^4$ lattice results.

The same can be said for the sigma susceptibility, which deserves,
however, a special remark.
A rather delicate point about the susceptibility computation is the
subraction of the disconnected part. In principle, this is not performed
by our noisy estimator .The long distance behaviour of the sigma propagator,
at least in the strong coupling limit, makes us confident that the extra
contribution is not significant in our case. Moreover, we can check our
data against the numerical derivative of the data for the
chiral condensate on the $24^4$ lattice, as shown in the EOS section.
Both strategies (inversion in a noisy background, numerical derivative)
are possibly objectionable, but for different reasons. Their agreement
gives us reasonable confidence in the results.

\subsection{Spectroscopy analysis}

{}~~~~The spectroscopy data are from a $16^3\times 32 $ lattice, where a
point, a wall source, and suitable point split wall sources were used
for the inversion of the Dirac propagator. We took some measurements
also on a $16^3\times 64$ lattice for checking purposes.
The computation of the spectrum
with the (rigid) point split wall requires gauge fixing. The gauge
is not fixed in the noisy wall inversion, where our aim is
to obtain the point propagator with better statistics. Thus, in this
case all the gauge dependent contributions have to cancel out.

\begin{table}
\caption{Results for   from the
effective mass analysis of the wall propagators in the pseudoscalar
channel
(t=14).}
 \begin {tabular} {|l|l l l l l |} \hline
     & .250    & .255    & .260    & .270    & .280  \\
  \hline
 .003& .1798(60) &.1727(88)&.1473(83)&.1040(111)&.0609(251)       \\
 .005& .2288(44) & --      & --      &.1521(70) &.1133(12)       \\
 .007& .2676(36) & --      & --      &.1927(52) &.1537(82)       \\
 .009& .3003(31) & .2925(41)&.2761(38)&.2283(42)&.1884(63)       \\
 .011& --        & --      & --      &.2620(45) & --    \\
 .013& --        & --      & --      &.2907(51) & --    \\
 .015& --        & .3708(29)&.3602(27)&.3173(46)& --    \\
 .017& --        & --      & --      &.3421(43) & --    \\
 .019& .4242(35) & --      & --      &.3665(48) & --    \\
 .021& .4423(34) & .4317(24)&.4247(21)&.3888(45)& --    \\
 .023& .4593(33) & --      & --      &.4096(43) & --    \\
 .025& .4754(32) & --      & --      &.4293(41) & --    \\
 \hline
 \end{tabular}
\label{tabpie1}
\end{table}

\begin{table}
\caption{Results from the
effective mass analysis of the point source
propagators in the pseudoscalar channel (t=14)}
 \begin {tabular} {|l|l l l l l |} \hline
     & .250    & .255    & .260    & .270    & .280  \\
  \hline
 .003& .1819(73)&.1764(108)&.1598(12)&.1267(102)&.0665(468)\\
 .005& .2346(59)&--&--&.1713(68)&.1174(202)\\
 .007& .2758(51)&--&--&.2112(53)&.1560(130)\\
 .009& .3098(43)&.3019(51)&.2932(52)&.2470(45)&.1906(98)\\
 .011& --&--&--&.2791(69)&--\\
 .013& --&--&--&.3088(60)&--\\
 .015& --&.3827(39)&.3768(39)&.3362(54)&--\\
 .017& --&--&--&.3615(49)&--\\
 .019& .4239(49)&--&--&.3909(46)&--\\
 .021&.4414(48)&.4430(33)&.4396(31)&.4132(44)&--\\
 .023&.4578(46)&--&--&.4340(42)&--\\
 .025&.4734(45)&--&--&.4536(40)&--\\
 \hline
 \end{tabular}
\label{tabpie2}
\end{table}

\begin{table}
\caption{ Results for the pion mass from two particle fits
for $t>2$ of the point/noisy
propagators}

 \begin {tabular} {|l|l l l l l |} \hline
     & .250    & .255    & .260    & .270    & .280  \\
  \hline
 .003&.1843(44)& .1500(323)&.1484(87)&.1145(64)&.0638(120)       \\
 .005&.2349(36)& --      & --      &.1591(55)  &.1149(74)       \\
 .007&.2752(27)& --      & --      &.2045(38)  &.1609(198)       \\
 .009&.3054(209)&.2923(112)&.2899(39)&.2419(38)&.1930(100)       \\
 .011& --      & --      & --      &.2748(67)  & --    \\
 .013& --      & --      & --      &.3099(68)  & --    \\
 .015& --      &.3679(228)&.3738(61)&.3372(66) & --    \\
 .017& --      & --      & --      & .3615(69) & --    \\
 .019&.3856(302)& --      & --      & .3853(71)& --    \\
 .021&.4037(280)&.4127(342)&.4191(512)&.4000(58)& --    \\
 .023&.4183(443)& --      & --      &.4195(99)  & --    \\
 .025&.4281(975)& --      & --      &.4375(3165)& --    \\
 \hline
 \end{tabular}
\label{tabpifi1}
\end{table}

\begin{table}
\caption{Results for the first excited state in the pseudoscalar
channel, from two particle fits for $t>2$ of the noisy propagators}
\label{tabpifi2}

 \begin {tabular} {|l|l l l l l |} \hline
     & .250    & .255    & .260    & .270    & .280  \\
  \hline
 .003& .5022 (1179) &.3364(79) &.4584(55)& .5561(317) & .5218(27)      \\
 .005& .5380 (920)  & --       & --      & .5642(232) & .5495(18)      \\
 .007& .5982 (870)  & --       & --      & .5798(208) & .5653(17)      \\
 .009& .5420 (44)   &.4980(722)&.6289(481)& .5962(199)& .5773(18)      \\
 .011& --           & --       & --      & .6730(486) & --    \\
 .013& --           & --       & --      & .6958(488) & --    \\
 .015& --           &.5246(1467)&.7364(349)&.7179(499)& --    \\
 .017& --           & --       & --      &.7405(510)  & --    \\
 .019& .4630 (283)  & --       & --      &.7057(433)  & --    \\
 .021& .4786 (196)  &.5292(593)&.5764(247)&.7261(.4521) & --    \\
 .023& .4917 (244)  & --       & --      &.7467(.47)    & --    \\
 .025& .4962 (461)  & --       & --      & .7613(1.17)  & --    \\
 \hline
 \end{tabular}
\end{table}

\begin{table}
\caption{Amplitude of the fundamental state from two particle fits for
$t> 2$ of $G_\pi(t)/G_\pi(0)$}

 \begin {tabular} {|l|l l l l l |} \hline
     & .250    & .255    & .260    & .270    & .280  \\
  \hline
 .003&.711(63) & .413(135) & .376(48)&.234(24)&.104(21)\\
 .005&.708(54) & --      & --      &.225(19) &.115(15) \\
 .007&.726(47) & --      & --      &.231(18) &.121(15) \\
 .009&.701(270)& .481(101) & .492(33)&.240(18) &.126(15)\\
 .011& --      & --      & --      &.268(40) & --    \\
 .013& --      & --      & --      &.288(40) & --    \\
 .015& --      & .474(254) & .534(54)&.309(44)& --    \\
 .017& --      & --      & --      & .333(41) & --    \\
 .019&.272(180)& --      & --      & .349(34) & --    \\
 .021&.34(31)  & .343(296) & .431(490)&.324(30) & --    \\
 .023&.30(17)  & --      & --      & .329(55)   & --    \\
 .025&.21(42)  & --      & --      & .34(93)   & --    \\
 \hline
 \end{tabular}
\label{taba1}
\end{table}

\begin{table}
\caption{Amplitude of the 1st excited state  from two particle fits for
$t> 2$ of $G_\pi(t)/G_\pi(0)$}
 \begin {tabular} {|l|l l l l l |} \hline
     & .250    & .255    & .260    & .270    & .280  \\
  \hline
 .003& .264 (41)&.549(192)& .539(57)& .671 (40)& .710(46)      \\
 .005& .253 (39)& --      & --      & .674(37) & .719(46)      \\
 .007& .231 (35)& --      & --      & .667(35) & .719(46)      \\
 .009& .227 (163)& .412(90)& .453(40)&.656(33) & .716(46)      \\
 .011& --      & --      & --      & .705(88)  & --    \\
 .013& --      & --      & --      & .689(85)  & --    \\
 .015& --      &.397(197) & .386(37)&.671(83)  & --    \\
 .017& --      & --      & --      & .651(80)  & --    \\
 .019& .643(256)& --      & --      & .622(52) & --    \\
 .021& .649(190)& .552(286) & .425(460)& .600(300) & --    \\
 .023& .675(248)& --      & --      & .578(359)    & --    \\
 .025& .829(517)& --      & --      & .562(880)    & --    \\
 \hline
 \end{tabular}
\label{taba2}
\end{table}

\begin{table}
\caption{ Sigma masses from fits for $t>3$ to $Aexp(-m_\sigma)+B$ of the noisy
scalar propagators}
\begin {tabular} {|l|l l l l l |} \hline
     & .250    & .255    & .260    & .270    & .280  \\
  \hline
 .003&.4566(245)&.5289(339)&.4706(183)&.5663(180)& .6152(274)       \\
 .005&.5095(189)& --      & --      &.5567(131) & .6006(167)      \\
 .007&.5926(197)& --      & --      &.5697(95)  & .6019(116)      \\
 .009&.5888(148)&.5817(161)&.5490(96)&.5951(98) & .6138(111)       \\
 .011& --      & --      & --      &.6034(138)  & --    \\
 .013& --      & --      & --      & .6144(194) & --    \\
 .015& --      & .6551(114)&.6452(85)&.6344(112)& --    \\
 .017& --      & --      & --      &.6462(130)  & --    \\
 .019&.7488(316)& --      & --      &.6658(120) & --    \\
 .021&.7264(209)& .7367(127)&.7076(87)&.7055(85)& --    \\
 .023&.7969(250)& --      & --      &.6994(84)  & --    \\
 .025&.8094(188)& --      & --      &.7337(108) & --    \\
 \hline
 \end{tabular}
\label{sigm}
\end{table}

\begin{table}
\caption{Background B from fits for $t>3$ to $Aexp(-m_\sigma)+B$ of the noisy
propagators in the scalar channel}
 \begin {tabular} {|l|l l l l l |} \hline
     & .250    & .255    & .260    & .270    & .280  \\
  \hline
.003&9.2(20.6)E-04&-2(189)E-05&7.00(1.76)E-03&1.33(18)E-02&1.40(27)E-02
\\
.005&12.3(5.7)E-04   & --  & --   &4.76(68)E-03&7.07(1.03)E-03       \\
 .007&1.4(2.5)E-04    & --             & -- &2.18(31)E-03&3.21(48)E-03       \\
.009&-1.8(14.8)E-05&8(1430)E-07&2.21(1.22)E-04&1.18(13)E-03&2.38(24)E-03
 \\
 .011& --              & --             & --           &5.38(1.66)E-04& --
\\
 .013& --              & --             & --           &3.96(93)E-04         &
--    \\
 .015& --              &2.00(3.42)E-05  &4.31(2.56)E-05&1.53(60)E-04         &
--    \\
 .017& --              & --             & --           &7.82(3.37)E-05
& --    \\
 .019&-2.5(2.2)E-05    & --             & --           &1.25(21)E-04         &
--    \\
 .021&2.17(1.49)E-05   &5.82(9.16)E-06  &2.97(0.88)E-05&6.58(1.68)E-05
& --    \\
 .023&1.07(0.99)E-06   & --             & --           &3.13(1.16)E-05
& --    \\
 .025&-7.58(7.74)E-06  & --             & --           &1.11(86)E-05         &
--    \\
 \hline
 \end{tabular}
\label{sigmab}
\end{table}

\begin{table}
\caption {Electron masses from point source propagators.}
 \begin {tabular} {|l|l l l l l |} \hline
     & .250    & .255    & .260    & .270    & .280  \\
  \hline
 .003& .187(30)&.157(45) &.113(38) &.094(11)&.063(14)       \\
 .005& .215(27)& --      & --      &.115(11)&.080(12)       \\
 .007& .232(26)& --      & --      &.133(12)&.097(12)       \\
 .009& .246(24)&.204(23) & .181(28)&.150(12)&.113(13)       \\
 .011& --      & --      & --      &.118(18)& --    \\
 .013& --      & --      & --      &.134(17)& --    \\
 .015& --      &.236(19) & .235(27)&.147(17)& --    \\
 .017& --      & --      & --      &.160(17)& --    \\
 .019& .293(69)& --      & --      &.198(16)& --    \\
 .021& .310(70)&.265(17) & .279(27)&.210(16)& --    \\
 .023& .325(70)& --      & --      &.223(17)& --    \\
 .025& .341(71)& --      & --      &.235(18)& --    \\
 \hline
 \end{tabular}
\label{femt}
\end{table}

\begin{table}
\caption{Results for the $\rho$ mass  from two particle fit of the wall
propagators.}
 \begin {tabular} {|l|l l l l l |} \hline
     & .250    & .255    & .260    & .270    & .280  \\
  \hline
 .003&.3394(94)&.2816(99) &.2293(58)&.1661(44) & .1206 (47)      \\
 .005&.3387(181)& --      & --      &.2097(34) & .1630(35)      \\
 .007&.4226(61) & --      & --      &.2476(29) & .1981(30)      \\
 .009&.4573(302) &.4107(509)&.3539(66)&.2821(26) & .2295(28)      \\
 .011& --      & --      & --      &.3126(44)         & --    \\
 .013& --      & --      & --      & .3414(37)        & --    \\
 .015& --      & .5016(131) &.4446(67)&.3681(116)         & --    \\
 .017& --      & --      & --      & .4000(403)        & --    \\
 .019&.6276(122)& --      & --      &.4202(286)         & --    \\
 .021&.6501(139)&.5757 (173)&.5188(69)&.4442(234)         & --    \\
 .023&.6717(138)& --      & --      & .4672(372)        & --    \\
 .025&.6937(148)& --      & --      & .4895(330)        & --    \\
 \hline
 \end{tabular}
\label{tabrho}
\end{table}

\begin{table}
\caption{Psudoscalar meson masses for $m_e = .003$.
Weighted results of effective masses, $t>5$}
 \begin {tabular} {|l|l l l l l |} \hline
     & .250    & .255    & .260    & .270    & .280  \\
  \hline
$\gamma_5\xi_5$
&.3353(477)  &  .2684(411) &   .2431(263)& .1698(159) & .1265(157) \\
$\gamma_5\xi_k\xi_5$
& .3266(226) &  .2803(195) &   .2336(125)& .1499(70)  & .0979(77)  \\
$\gamma_4\gamma_5\xi_l\xi_m$
& .3469(338) &  .2677(268) &  .2394(170) & .1678(103) & .1324(111) \\
$\gamma_4\gamma_5\xi_m$
& .3238(208) &  .2951(201) & .2233(108)  & .1535(65)  & .1011(75)  \\
$\gamma_5\xi_m\xi_4$
& .3457(293) &  .2884(260) & .2461(149)  & .1636(87)  & .1346(91)  \\
$\gamma_5\xi_4$
& .3274(194)  &  .2735(185) & .2305(105)  & .1537(63)  & .1072(74)  \\
 \hline
 \end{tabular}
\label{tabm03}
\end{table}

\begin{table}
\caption{Psudoscalar meson masses for $m_e = .009$.
Weighted results of effective masses, $t>5$}
 \begin {tabular} {|l|l l l l l |} \hline
     & .250    & .255    & .260    & .270    & .280  \\
  \hline
$\gamma_5\xi_5$
&.4544(287)&.4103(256) & .3628(156) &.2813(92) &.2319(83)       \\
$\gamma_5\xi_k\xi_5$
&.4522(141)&.4071(128) & .3605(85) &.2744(48)  & .2168(48)       \\
$\gamma_4\gamma_5\xi_l\xi_m$
&.4629(181)&.4022(193) &  .3645(110) &.2815(67) & .2330(69)       \\
$\gamma_4\gamma_5\xi_m$
&.4568(135)&.4143(138) & .3583(82) &.2762(46) & .2216(44)      \\
$\gamma_5\xi_m\xi_4$
&.4644(167)&.4073(164) & .3656(99) &.2822(58)  & .2369(58)    \\
$\gamma_5\xi_4$
&.4521(118)&.4086(134) &.3581(74) & .2763(47)  & .2334(49)     \\
 \hline
 \end{tabular}

\label{tabm09}
\end{table}

For the meson operators the standard parametrization of the
propagators
reads:

\begin{eqnarray}
G(\tau) & = & a [exp(-M\tau) + exp(-M(N_\tau - \tau)] + \\ \nonumber
& & a1 [exp(-M1\tau) + exp(-M1(N_\tau - \tau)] + \\ \nonumber
& & (-1)^{\tau}\tilde a[exp(-\tilde M\tau)+exp(-\tilde M(N_{\tau}-\tau)
\end{eqnarray}

In many cases, the oscillating channel, and/or the radial excitation
proved to be irrelevant. For the sigma particle, the inclusion of
a constant (background) term was necessary, as discussed below.

For the pion, the contribution of the oscillating channel is very
small in the strong coupling phase.
At weak coupling, especially with the wall operator, there is a modest
oscillation which can be dealt with by extracting
the direct channel contribution
in the usual way. We have thus performed the effective masses analysis
of the direct channel, computed as
\begin{equation}
G_{direct}(2t) = 2G(2t) - G(2t-1) -G(2t+1)
\end{equation}
(note that according to ~\cite{GSST84} this should give us the exact propagator
in the direct channel).

Many of the results are displayed
in Fig.14--16. We observed some discrepancies, at the
smaller values of the quark masses,
between the results with the point and the wall source.
This occurred even for the apparently asymptotic parts of the propagators.
In these cases we
preferred the results of two particle fits (M and M1 according
to eq. (18))
(the inclusion of an oscillating state $\tilde M$ proved to be
irrelevant).
The mass
of the fundamental
particle, the Goldstone boson in the strong coupling region, is
in between the results for the effective masses from point and wall
sources, as it should be, since these provide lower and upper bounds,
respectively.  We show
a sample of the results of the fits in Fig.~17, whose $\chi^2$'s are in the
range
of a few percent, very small as they should be for correlated data.

Rather than discussing the systematics further, we prefer to show
the full collection of the results
in Tables~ \ref{tabpie1}, ~\ref{tabpie2}, ~\ref {tabpifi1}.
The errors come from jack-knife estimates, performed by discarding
one propagator at a time. Wherever they agree, the effective mass
result, either from a wall or point source
measurement, is more accurate. In case of (mild) disagreement, the fit
result is safe.
We show also the results of the first excited state, again from the
two particle fits, in Table~\ref{tabpifi2}.

The behaviour of the $\sigma$ particle is rather peculiar. Its main
characteristic is a plateau in the propagators which shows up in
the weak coupling region.

In the strong coupling region, one particle
fits were compelling, and in agreement with effective mass analyses.
(in this case the oscillating component is small, so we could
use the standard even-even, odd-odd combinations for the extraction
of the mass values). In the weak coupling region, a one-particle
fit badly fails.

Two particle (M and M1) fits give results for
the lower mass
in agreement with one-particle fits in the strong coupling region,
although noisier, while in the weak coupling region, at small quark
masses, the fundamental mass
is extremely low, often compatible with zero.

A very reasonable ansatz which works in the whole range of parameters
was thus to fit the propagators with a
form  $A exp^{-mt} + B$, with symmetrization (Figs.~18 and 19).
$B$ is compatible with
zero in the strong coupling region, with masses in agreement with the
ones estimated via the one particle fit.
On the other hand, in the weak coupling
region, and at small masses, $B$ is distinctly different from zero, and the
mass is consistent with the one determined via a two particle fit.  Not
surprisingly,
the situation at $\beta = .260$ is the most ambiguous, and it is not
clear if there is a background, a real pole mass, or what.
For instance, we show in Fig. 20 the results from a fit with background
for $t>3$(dot), and two particle fit for $t>2$ (dash) at $\beta = .260$
and $m = .003$. Both $\chi^2$ are a few percent, and the fits
are practically indistinguishable. The mass for the fit
with background is .47(2) and the background itself is 7(1)E-03.
The two particle fit gives a mass of .51 and a smaller mass
$.13$ with an amplitude of 6E-02.

 It is unlikely that a low mass particle really exists here
since the background, or, equivalently, its
amplitude, decreases when the bare quark mass increases.
The power law behaviour,
discussed in ~\cite{HKK93}, could be appropriate in the immediate
vicinity of the critical region, but a detailed test of this hypothesis
would require much more statistics, insight etc.
Also, recall that the naive symmetrization used here is not compelling.
Multiple images can alter the behavior
on a torus from that expected on an infinite volume.
This is true even for the naive exponential behavior, and probably even
more for some slow-decaying function.  Apparently, a fixed background
is enough to summarize what is going on, thus
for the time being, we take our results from  fits
in the form discussed above. We  quote
the results, with the extra-caveats
discussed above  for  $\beta = .260$, in Tables ~\ref{sigm},
{}~\ref{sigmab}.

A rather convincing argument in favour of the chosen parametrization
comes from the study of effective masses once the background $B$
was subracted `by hand'. $B$ can be estimated from the long distance
behaviour of the sigma propagator, once the plateau has set in
($t=13$ proved to be a reasonable starting point for the computation
of the background).
We show the results in Fig.~21: the effective masses computed
after the subtraction of the background
flatten, giving estimates in agreement with the ones
obtained with the background fits.  We also show
what we would have obtained without subraction, a decreasing
plot suggesting an unnaturally low mass.

To learn more about the nature of the background, we measured the
propagators on a bigger lattice, for the four smaller masses at $\beta
= .270$. We see (Fig. 22) that the background
is clearly reduced, and we tentatively
conclude
that it becomes less and less significant while increasing the
lattice length.
Our parmetrization explains the
odd behaviour of the $\sigma$ susceptibility on `small' cylindrical
lattices, and it explains why the symmetric lattices are apparently
satisfactory (they do not take information
from the unsafe long distance behavior,
dominated by finite size effects), and suggests that the
proper behaviour is recovered in the infinite volume limit.
We conclude that we have safe estimates of the mass in the sigma
channel which is dominating in the time distance 4--13, in lattice units.
As is evident from the plots, the signals are lost at larger distance,
and we cannot exclude that lower mass excitations are present in the
scalar spectrum. This, as discussed above, would be especially
relevant
when comparing the sigma mass with twice the mass of the fermion.

The fermion propagators is well described by the usual form
\begin{equation}
G(t) = a(e^{-mt} - (-1)^te^{-m(T-t)})
\end{equation}
for all
euclidean times. We have checked that the results do not change,
within rather large statistical errors, while changing the fitting interval.
Wherever the comparison is possible,
our new results obtained with a point source agree with the previous ones,
coming from wall source measurements.
The results of the fits (a subset of the plots
are shown in Fig. 23) are given in Table~\ref{femt}.

The $\rho$ particle does not couple appreciably to the point source,
so we quote here the results obtained with the wall source.
Wherever a cross check was possible, the two approaches
gave consistent results. We note that the fits look good
(a subset of them are shown in Fig. 24) but their $\chi^2$'s are not
very low (they are typically order 1, too high for correlated data).
The effective mass analysis, especially at strong coupling,
is not quite flat. Anyway,  it oscillates  around the results of the fit.
The new data deep in the symmetric phase confirm
that an oscillation with a very low ($\simeq 0$) mass is needed in order to
obtain good fits. As discussed for the sigma propagator, this very low mass
can be traded for a constant (in this case purely oscillating) term.
However, in this case the problem is less severe, since it does not affect
our estimate in the fundamental channel.

\section{Conclusions}

{}~~~~The results presented here include a detailed comparison
of the data with an equation of state \`a la Nambu Jona Lasinio,
together with a further study of the powerlaw scaling predicted
by analytic computations.
Our analysis favours
power law scaling with non-mean field exponents
in agreement with the ones found in past work.
We have performed detailed fits of old data for the two
Equations of State, and we have confirmed
their results by a new numerical analysis.
We have analyzed our new data for the susceptibilities, and contrasted
them with the predictions of the two scenarios. From any point of
view, the power--law hypothesis provides an accurate description of
the data, while the NJL analysis gives poorer results (in the case of
direct fits of the order parameter), or fails qualitatively (in the
case of the analysis of susceptibility data).
We stress that the finite size effects and other sources of systematic
errors were kept under control, as discussed in detail in the text.
We have also commented on possible corrections to the equation of
state, and  on their impact on the results. We found these corrections
to be small, at least in the case we were considering. This point
deserves however further investigation.

We have gained a good understainding of the Goldstone mechanism, and
of chiral symmetry breaking in the spectrum. We have found that
the candidate Goldstone boson decouples in the weak coupling phase.
The behaviour of the amplitudes in the (would be) Goldstone and
non-Goldstone mode is clear, and picks up the critical point
with good accuracy. This result also
 explains how the low-lying state in the pseudoscalar channel
mantains the property of a Goldstone boson -- a puzzling result of
previous works. We have analyzed in detail the behaviour of the
propagators in the scalar and vector sector, and found clear
indications of chiral symmetry restoration at the transition point.
A point left open is the level ordering $\sigma$ -- fermion. The
fermion mass was measured again, and found in good agreement with
previous work.  We have also obtained good estimates for the scalar
propagator  in the intermediate time distance region. However, the
long distance region still evades us, and we do not have firm
conclusions for the physically relevant low-lying pole of the scalar
propagator.
We mention that the source used for the inversion of the Dirac
operator is crucial in this case: in previous work with a wall source
we did not get any signal for the sigma. A computation of the sigma
mass probably requires a dedicated source (i.e. with the appropriate
quantum numbers).

Finally, we have presented the first computation in QED of the full
mesonic spectrum local in time. We have observed indications of
flavor/rotational symmetry restoration. However, it is not clear if
such behaviour is induced by the chiral transition, or by the
perturbative limit.

\vskip 1.0truecm

We would like to thank the National Science Foudation for support
under grant NSF-PHY92-00148. The computer simulations reported here
were done on the
 C90-CRAYs  at the Pittsburgh Supercomputer Center (PSC)
and the National Energy Research Supercomputer Center (NERSC),
and on the CM--2 at the PSC.
We wish
to thank Jean--Fran\c cois Laga\"{e} for conversations and Donald K.
Sinclair for some of the meson spectroscopy codes.

\newpage
\noindent
{\bf Figure captions}

\begin{enumerate}

\item
Results of power--law fits  for
various $\beta$ intervals.Circles, $.240 \le \beta \le .260$,
crosses, $.245 \le \beta \le .260$,
squares, $.250 \le \beta \le .260$.

\item
Results of the fits \`a la Nambu--Jona-Lasinio for
various $\beta$ intervals. Crosses, $.245 \le \beta \le .260$,
squares, $.250 \le \beta \le .260$.

\item
$R_t$ vs $R_m$. The data are from the $24^4$ lattice, the logarithmic
derivatives are evaluated numerically. The input value for the critical
$\beta$ was .2572. The straight line is the prediction of the EOS fit,
$R_t = (1 - \delta R_m)/ (\delta - 1)$.

\item
(a)Susceptibility ratio , $16^4$ lattice, circles. $\beta$ = .250, .255, .260,
.265, .270, .275, .280, from bottom to top. The crosses are for the
same ratio, evaluated by taking the logarithmic derivative numerically on
the $24^4$ lattice, at $\beta$ = .250, .255, .260. The straight line is drawn
at $1/\delta$, with $\delta = 2.12$, and shows that $ .255 < \beta_c < .260 $
in agreement with the results of the EOS fit.
(b)Zoomed view around the critical region. Also, the $24^4$
lattice data are shown in steps of $\Delta \beta = .001$, beginning
from $\beta = .260$ from top to bottom.

\item
$\chi_\pi/\chi_\sigma - 3$ vs $1/\log <\bar\psi\psi>$. Data in the $\beta$
interval $.250 - .260$, $\Delta \beta (.005,.001)$, (circles, crosses),
($16^4$, $24^4$). The solid line is the NJL prediction at criticality,
the dashed one is the power law prediction,
consistent with the data at $\beta =
.257$.

\item
Fermion mass
as a function of $<\bar\psi\psi>$. The data for the chiral condensate
are from $24^4$ lattice, the data for the fermion mass are taken
on a $16^3 \times 32$ lattice, ref. [5]. The solid line is the free
fermion prediction, the dashed one the strong coupling limit. $m_e$ =
(.002,.003,.004,.005),  (diamonds, crosses, squares, circles).

\item
Amplitude of the Goldstone mode (solid), and sum of the amplitudes
of the excited states (dash) for $m_e = .003$ (top) and
$ m_e = .009$ (bottom),
as a function of $\beta$. The Goldstone mode becomes dominant around
the critical $\beta$, the effect being more pronounced at small mass.

\item
Ratio of the sigma to the pion propagator at zero distance. Squares,
circles, crosses, fancy crosses are for $m_e$ = .003, .009, .015, .021.
The short distance behaviour shows clear indications of chiral symmetry
breaking/restoration at small masses.

\item
Ratio of the A1 to the $\rho$ propagator at zero distance. Squares,
circles, crosses, fancy crosses are for $m_e = .003, .009, .015, .021$.
Also in this case, the short distance behaviour shows clear indications
of chiral symmetry  breaking/restoration at small masses.

\item
Effective mass plot for the pseudoscalar mesons at
$\beta = .260, m_e = .003$.
The Goldstone is the lowest, all the others are degenerate.

\item
Mass of the Goldstone (diamonds) and of the
$\gamma_5\xi_k\xi_5$ (crosses)
mesons as a function of $\beta$. $m_e = .003$, upper portion of the figure,
$m_e = .009$ lower.

\item
Finite size effects on the chiral condensate in the strong
coupling region at $\beta = (.250, .255)$ (solid, dashed). Squares
are for the $24^4$ lattice, diamonds for the $16^4$, crosses for
$8^4$.

\item
Pion (dash) and sigma (solid) susceptibilities at $\beta = .280$.
Crosses, diamonds, squares are for the $8^4$, $16^4$, $16^3 \times 32$
lattices.

\item
Effective mass  for the pion at quark masses $.003$ (crosses) and
$.009$  (circles) at $\beta = .250$, point and wall source.

\item
 Effective mass plot for the pion at $\beta = .250$, $m_e = .019$ (crosses).
The results from an high-statistics
run on a $16^3 \times 64$  lattice are shown for comparison (circles).

\item
Effective mass  for the pion at quark masses $.003$ (crosses) and
$.009$  (circles) at $\beta = .280$, point and wall source.

\item
Two particle fits for the pseudoscalar propagator (point source)
at $\beta = .260$, masses .003, .009, .051, .021, $t>2$. The $\chi^2$
ranges from 2/1000 to 6/100.

\item
Scalar propagator (noisy source) fits to $Ae^{-m_\sigma t} +B$, $m_e = .003$,
$t>3$. Crosses,
diamonds, squares are for $\beta = .250, .260, .280$. The $\chi^2$'s are
5,7,9 /100.

\item
Scalar propagator (noisy source) fits to $Ae^{-m_\sigma t} +B$, $m_e = .007$,
$t>3$. Crosses,
diamonds, squares are for $\beta = .250, .270, .280$. The $\chi^2$'s are
3,40,30 /100.

\item
Scalar propagator (noisy source) fits at $\beta = .260$ and $m_e = .003$.
The results of a fit with background for $t> 3$ (dot) and
a two particle fit for $t>2$ (dash) are practically coincident.

\item
Effective mass (even-even, odd-odd) for the sigma particle.
{}From top to bottom, $(\beta = .250, m_e = .007), (\beta = .260, m_e = .003),
(\beta = .280, m_e = .007)$. The squares are the naive results,
the crosses with the subtracted background ($t>13$).

\item
Scalar propagator at $\beta = .270, m_e = .003$ on the $16^3 \times 32$ and
$16^3 \times 64$ lattices. Note the agreement at short distance, and the
clear size-dependent plateau showing up at large time separations.

\item
 Electron propagator fits in the interval $0 < t < 17$, $m_e = .009$,
$\beta = .250,.255, .270, .280$ from bottom to top.


\item
Two particle fits  (direct + oscillating) for $t>4$ to the vector propagator
at $m_e = .003$, $\beta$ ranges from $.250$ (bottom) to $.280$ (top).

\end{enumerate}
\end{document}